# A parallel implementation of a derivative pricing model incorporating SABR calibration and probability lookup tables[☆]

Qasim Nasar-Ullah[1]

*University College London, Gower Street, London, United Kingdom*

## Abstract

We describe a high performance parallel implementation of a derivative pricing model, within which we introduce a new parallel method for the calibration of the industry standard SABR (stochastic-$\alpha\beta\rho$) stochastic volatility model using three strike inputs. SABR calibration involves a non-linear three dimensional minimisation and parallelisation is achieved by incorporating several assumptions unique to the SABR class of models. Our calibration method is based on principles of surface intersection, guarantees convergence to a unique solution and operates by iteratively refining a two dimensional grid with local mesh refinement. As part of our pricing model we additionally present a fast parallel iterative algorithm for the creation of dynamically sized cumulative probability lookup tables that are able to cap maximum estimated linear interpolation error. We optimise performance for probability distributions that exhibit clustering of linear interpolation error. We also make an empirical assessment of error propagation through our pricing model as a result of changes in accuracy parameters within the pricing model's multiple algorithmic steps. Algorithms are implemented on a GPU (graphics processing unit) using Nvidia's Fermi architecture. The pricing model targets the evaluation of spread options using copula methods, however the presented algorithms can be applied to a wider class of financial instruments.

## 1. Introduction

The pricing of financial derivatives is a computationally demanding task and for many users of financial derivatives such pricing is undertaken by large scale computing farms. The use of highly parallel co-processors such as GPUs (graphics processing units), as opposed to the sole use of CPUs (central processing units), is thought to offer computational speed and cost improvements offset against the complexity of deriving algorithms able to substantially exploit highly parallel architectures. Computational speed and cost improvements are based on the observation that highly parallel architectures such as GPUs typically exhibit superior levels of peak memory and arithmetic throughput rates compared to CPU architectures.

Within the context of financial derivatives pricing, a diverse set of computational techniques exist, with traditional methods constituting the following: *analytic methods* which calculate closed-form solutions of partial differential equations (PDEs) (Black and Scholes, 1973) or probabilistic expectations (Cox and Ross, 1976), *finite difference methods* (Schwartz, 1977) which approximate solutions to PDEs using difference equations, *Monte Carlo methods* (Boyle, 1977) which attempt to simulate stochastic processes and finally, *tree based methods* (Cox et al., 1979) which create lattice structures for stochastic processes. Aside from analytic methods, which do not require significant computational effort, several works have demonstrated how the inherent parallelism exhibited by such computational methods enables effective use of highly parallel co-processors such as GPUs. In particular, Monte Carlo methods, due to their often significant levels of inherent parallelism, have shown particular performance gains over CPU based alternatives (Joshi, 2010; Bennemann et al., 2008; Dixon et al., 2012). Modern pricing methods for financial derivatives are known to rely on additional computational techniques, such as iterative minimisation algorithms, which are often computationally expensive (Brigo and Mercurio, 2006). As such, within this paper we formulate high performance parallel algorithms for the acceleration of a derivative pricing model that is composed of both Monte Carlo and iterative algorithms, where we focus solely on iterative algorithms

---

[☆]Under review: Updated January 13, 2013
[1]*Email address:* q.nasar-ullah@ucl.ac.uk

operating in the presence of variable accuracy parameters.

The first type of iterative algorithm we consider is the calibration of the SABR (stochastic-$\alpha\beta\rho$) model (Hagan et al., 2002). We present a new parallel method which is designed to be implemented on highly parallel architectures such as GPUs. SABR calibration is used to ensure that the output from the SABR model agrees with market observations of implied volatility, this essentially corresponds to a minimisation problem. SABR models are common within derivative pricing and although it is typical for financial market participants to use customised forms of the SABR model, the method we present is applicable to all SABR models consistent with a set of generic properties we highlight within the paper. Typically the SABR model calibration is through minimisation algorithms optimised and designed for single thread execution, as such known methods to calibrate SABR models (West, 2005; Clark, 2011; Nilsson, 2008) include off-the-shelf Simplex methods (Nelder and Mead, 1965), Levenberg-Marquardt methods (Marquardt, 1963) and gradient descent methods (Press et al., 2007). Our new parallel SABR calibration method operates by iteratively refining a two dimensional grid until a unique set of SABR parameters is found that match SABR generated option volatilities to a standard input of three market observed option implied volatilities. The method is based on principles of surface intersection and is shown to offer guaranteed convergence.

The second type of iterative algorithm we consider is the creation of lookup tables used to store cumulative probability distributions. We present GPU based parallel algorithms that create dynamically sized lookup tables, are able to cap maximum estimated linear interpolation error and are optimised for probability distributions of the type implied by the SABR model which exhibit linear interpolation error clustering.

Our two stated iterative algorithms are combined within a derivative pricing model, for which we empirically highlight error propagation and performance results. We note that in addition to the possibility of using parallel algorithms for a single SABR calibration or the creation of a single lookup table, further parallelism is exhibited by our pricing model as it contains multiple independent coupons, each of which require the same calculations (Nasar-Ullah, 2012). Results are generated using the system properties shown in Appendix A.

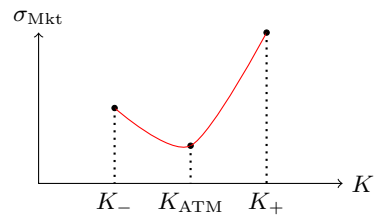

Figure 1: An example of the market implied volatility $\sigma_{\text{Mkt}}$ of three European options exhibiting a volatility smile. All three options contracts are identical except that they have different strike prices $K$.

*1.1. Structure of paper*

In section 2 we introduce a new parallel method for SABR calibration. In section 3 we introduce parallel algorithms for the creation of cumulative probability lookup tables. In section 4 we turn to our derivative pricing model as a whole (comprising of both SABR calibration and the creation of cumulative probability lookup tables), where we first empirically highlight error propagation through the pricing model as a result of changing accuracy parameters and secondly present overall timing results.

## 2. SABR model calibration

*2.1. Background*

Options are financial contracts where at an initial time $t = 0$ the buyer of the option acquires the right but not the obligation to fulfil a given transaction at a later time. The simplest type of option, the European call option, confers upon the holder the right to buy an underlying asset $F$ at a future time $T$ for a fixed strike price $K$. The payoff (i.e. the final value of the option at time $T$) can be expressed as follows, where $F(T)$ is the final asset price:

$$\text{Call payoff} = \text{Max}(F(T) - K, 0). \tag{1}$$

The seminal Black Scholes formula (Black and Scholes, 1973) is used to give the initial price $V$ of such options at $t = 0$. The Black Scholes formula is a function $V(F(0), K, T, r, \sigma)$, where additionally $F(0)$ is the initial asset price, $r$ is a risk free rate and $\sigma$ is a measure of asset volatility.

The market implied volatility $\sigma_{\text{Mkt}}$ is defined as the volatility such that when input into the Black Scholes formula, alongside the parameters $F(0), K, T, r$ of an observed market option with price $V_{Mkt}$, the resulting Black Scholes price matches $V_{Mkt}$, that is $V_{Mkt} = V(F(0), K, T, r, \sigma_{Mkt})$. Empirical evidence for several markets suggests the presence of a 'volatility smile' (Hagan et al., 2002), a phenomenon where identical option contracts except for different strikes $K$ have different implied volatilities as shown in figure 1. The Black Scholes model however assumes a constant volatility when calculating the price of such options, in doing so a single set of Black Scholes parameters cannot quantify a volatility smile.

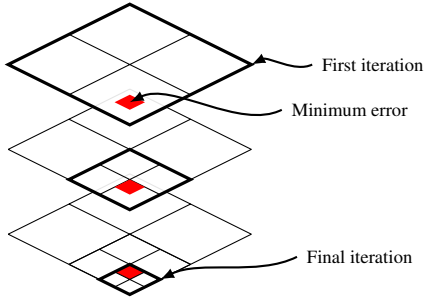

Figure 2: An example of a 2 × 2 grid refining over a two dimensional space. The grid iterates from coarse spacing to fine spacing in order to identify the precise location of the global minimum error. The global minimum error is located in the filled region.

With a view to capturing volatility smile dynamics, the SABR stochastic volatility model (Hagan et al., 2002) is a popular choice which (unlike the Black Scholes model) is able to generate strike dependent volatilities $\sigma_{SABR}$. Several SABR parameters must be estimated to fully specify a SABR model. The estimation of SABR parameters is typically conducted by a three dimensional minimisation, referred to as calibration, that ensures strike dependent SABR volatilities $\sigma_{SABR}$ replicate actual market implied volatilities $\sigma_{Mkt}$ for a number of market observed options.

The general SABR model can be given as:

$$\begin{aligned}
dF(t) &= s(t)F(t)^\beta dW(t),\\
ds(t) &= \alpha s(t)dZ(t),\\
dW(t)dZ(t) &= \rho dt, \quad (2)
\end{aligned}$$

where $F(t)$ is an asset price process, $s(t)$ is a volatility process and $dW(t), dZ(t)$ are two stochastic Brownian motion processes. Additionally, the SABR model has five parameters: $\beta$ is an asset price exponent term, $\alpha$ is a term representing the volatility of the volatility process $s(t)$, $\rho$ represents a correlation between $dW(t)$ $dZ(t)$, $F(0)$ is an initial asset price which is directly observable and $s(0)$ is an initial volatility value.

Since the value of $F(0)$ is directly observable, there are four unknown remaining parameters which require estimation. In our implementation, SABR parameters attempt to replicate an observed market volatility smile based on a standard approach of observing the implied volatilities of *three* market options with strikes $K_-, K_{ATM}$ and $K_+$ (see figure 1), where $K_{ATM}$ refers to the option's strike price being equal to the initial price of the underlying asset (i.e. $K = F(0)$) which is known as an ATM (at-the-money) option, and where $K_- < K_{ATM} < K_+$. Corresponding market implied volatilities are given as $\sigma_{Mkt}(K_-), \sigma_{Mkt}(K_{ATM})$ and $\sigma_{Mkt}(K_+)$. Further, we use the approach of arbitrarily selecting a constant value for the parameter $\beta$ (where $0 \leq \beta \leq 1$) (Hagan et al., 2002; West, 2005). Consequently we require calibration of the three remaining parameters: $\alpha$, $\rho$ and $s(0)$. It is further noted that the value of $s(0)$ can be calibrated directly from fixed values of $\alpha$ and $\rho$ based on $\sigma_{Mkt}(K_{ATM})$ (Hagan et al., 2002; West, 2005).

The minimisation or calibration objective can be stated formally as follows, where SABR volatility $\sigma_{SABR}$ is represented by a function of the previously stated varying parameters as $\sigma_{SABR}(\mathbf{K}_i, \alpha, \rho, s(0))$, where $\mathbf{K} \in \{K_-, K_{ATM}, K_+\}$, $i = 3$ and where calibrated values are listed as $\alpha^*, \rho^*, s(0)^*$:

$$\{\alpha^*, \rho^*, s(0)^*\} =$$
$$\arg\min_{\alpha, \rho, s(0)} \sum_{i=1}^{3} |\sigma_{SABR}(\mathbf{K}_i, \alpha, \rho, s(0)) - \sigma_{Mkt}(\mathbf{K}_i)|. \quad (3)$$

Using a three strike calibration approach, the calibrated values $\{\alpha^*, \rho^*, s(0)^*\}$ uniquely minimise (3) such that the summation on the RHS (right hand side) of (3) becomes zero.

*2.2. A parallel method for SABR calibration*

Our proposed calibration method attempts to ascertain the location of the global minimum satisfying the minimisation in (3), i.e. the values $\{\alpha^*, \rho^*, s(0)^*\}$, by iteratively refining a two dimensional grid until a desired level of SABR calibration error is achieved, where SABR calibration error is calculated as the summation on the RHS of (3). An example of such iterative grid refinement can be seen in figure 2. An illustration of typical SABR calibration error based on changes to $\alpha$ and $\rho$ is shown in figure 3 (the parameter $s(0)$ has been previously calibrated for each pair of $\alpha$ and $\rho$). Within our proposed calibration method each iteration of the refined grid contains the global minimum, however (as seen in figure 3) the SABR error surface is non-linear, hence establishing the presence of a global minimum within arbitrary points of a grid is non-trivial. To overcome this problem our method relies on exploiting known characteristics of the underlying SABR model to form a method based on principles of surface intersection.

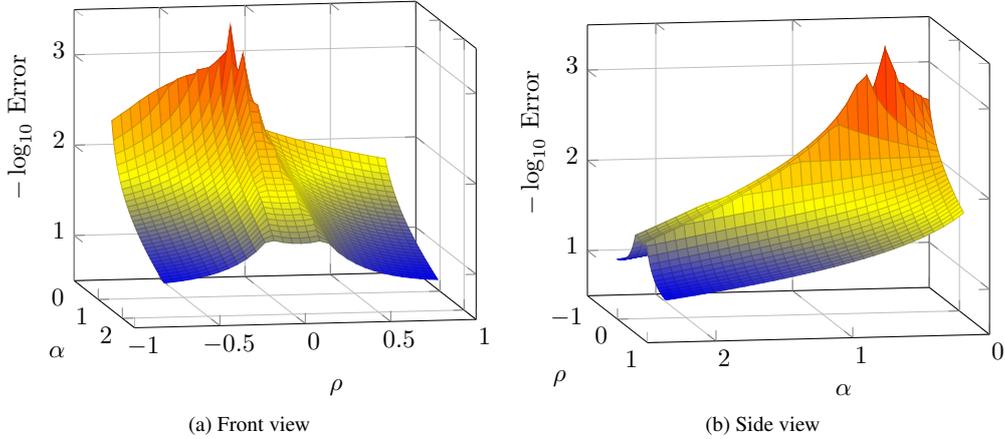

(a) Front view      (b) Side view

Figure 3: Typical SABR calibration error as calculated by the summation on the RHS of (3) where the parameters $\alpha$ and $\rho$ are varying. $s(0)$ has been previously calibrated based on (4) for each pair of $\alpha$ and $\rho$. Error is reported as $-\log_{10}$ error (e.g. 8.5 represents an error of $10^{-8.5}$) and the minimum error is located at the peaks.

We note that it is typical for separate market participants to use mathematically varied implementations of the original SABR model shown in (2). As a result our proposed calibration method is designed to be applicable to all SABR models consistent with a set of generic properties we highlight in the remainder of this section.

### 2.3. Preliminaries

As stated the parameter $s(0)$ can be calibrated from fixed values of the parameters $\alpha$ and $\rho$ based on $\sigma_{\text{Mkt}}(K_{\text{ATM}})$, thus we define $s(0)^*\langle\alpha,\rho\rangle$ as a value of $s(0)$ which is calibrated for a particular combination of $(\alpha,\rho)$ where:

$$s(0)^*\langle\alpha,\rho\rangle = \arg\min_{s(0)} |\sigma_{\text{SABR}}(K_{\text{ATM}},\alpha,\rho,s(0)) - \sigma_{\text{Mkt}}(K_{\text{ATM}})|. \quad (4)$$

Next we define error matrices $\mathbf{M}_-$, $\mathbf{M}_{\text{ATM}}$ and $\mathbf{M}_+$ which are of size $m \times n$, where $m$ is equal to the size of an array $\bar{\alpha}$ containing a range of $\alpha$ values indexed as $\{\alpha_1,\cdots,\alpha_m\}$, and $n$ is equal to the size of an array $\bar{\rho}$ containing a range of $\rho$ values indexed as $\{\rho_1,\cdots,\rho_n\}$. The matrices $\mathbf{M}_{(\cdot)}$ thus cover a two dimensional domain of $\alpha$ and $\rho$ with elements:

$$\mathbf{M}_{(\cdot)} = \begin{pmatrix} \mathbf{M}_{(\cdot)}^{\alpha_1,\rho_1} & \mathbf{M}_{(\cdot)}^{\alpha_1,\rho_2} & \cdots & \mathbf{M}_{(\cdot)}^{\alpha_1,\rho_n} \\ \mathbf{M}_{(\cdot)}^{\alpha_2,\rho_1} & \mathbf{M}_{(\cdot)}^{\alpha_2,\rho_2} & \cdots & \mathbf{M}_{(\cdot)}^{\alpha_2,\rho_n} \\ \vdots & \vdots & \ddots & \vdots \\ \mathbf{M}_{(\cdot)}^{\alpha_m,\rho_1} & \mathbf{M}_{(\cdot)}^{\alpha_m,\rho_2} & \cdots & \mathbf{M}_{(\cdot)}^{\alpha_m,\rho_n} \end{pmatrix}, \quad (5)$$

where each element $\mathbf{M}_{(\cdot)}^{\alpha,\rho}$ represents the error defined as:

$$\mathbf{M}_{(\cdot)}^{\alpha,\rho} = \sigma_{\text{SABR}}(K_{(\cdot)},\alpha,\rho,s(0)^*\langle\alpha,\rho\rangle) - \sigma_{\text{Mkt}}(K_{(\cdot)}). \quad (6)$$

Due to the minimisation within (4), each element within the matrix $\mathbf{M}_{\text{ATM}}$ has a value of zero (or a value representing the accuracy to which $s(0)$ was minimised within (4)). An example of the surfaces of $\mathbf{M}_-$ and $\mathbf{M}_+$ can be seen in figures 4a and 4b.

In the two dimensional domain of $\alpha$ and $\rho$ (within our error matrices $\mathbf{M}_-,\mathbf{M}_+$) we assume there is a point $C$ that represents the unique combination $(\alpha^*,\rho^*)$ where

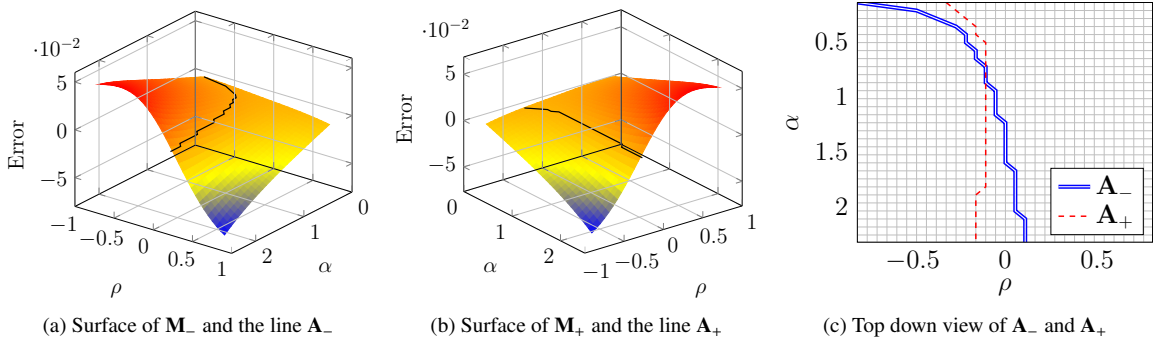

(a) Surface of $\mathbf{M}_-$ and the line $\mathbf{A}_-$  (b) Surface of $\mathbf{M}_+$ and the line $\mathbf{A}_+$  (c) Top down view of $\mathbf{A}_-$ and $\mathbf{A}_+$ both $\mathbf{M}_-^{\alpha^*,\rho^*} = 0$ and $\mathbf{M}_+^{\alpha^*,\rho^*} = 0$. Thus, the point $C$ represents the solution which uniquely satisfies the minimisation in (3). We note that based on (3), $\alpha$ and $\rho$ can be calibrated equivalently as follows, where $s(0)$ is implicitly calibrated due to (6):

$$\{\alpha^*, \rho^*, s(0)^*\} = \arg\min_{\alpha,\rho} |\mathbf{M}_-^{\alpha,\rho}| + |\mathbf{M}_+^{\alpha,\rho}|. \qquad (7)$$

Figure 4: (a) and (b) are typical examples of the surfaces of $\mathbf{M}_-$ and $\mathbf{M}_+$. Intersection of the surfaces of $\mathbf{M}_-$ and $\mathbf{M}_+$ with the plane $z = 0$ is shown by the solid lines defined as $\mathbf{A}_-$ and $\mathbf{A}_+$ respectively. The $z$ axis represents an error value shown in (6) stored within each element of our error matrices $\mathbf{M}_-$ and $\mathbf{M}_+$. (c) shows a top-down view of the intersection between $\mathbf{A}_-$ and $\mathbf{A}_+$, where the left hand side of $\mathbf{M}_-$ is positive upto $\mathbf{A}_-$ and the right hand side of $\mathbf{M}_+$ is positive upto $\mathbf{A}_+$.

Our proposed calibration method is based on the following generic properties, which are typical of all SABR models:

**Property 1.**

$$\frac{\partial \mathbf{M}_-}{\partial \rho} < 0, \qquad \frac{\partial \mathbf{M}_+}{\partial \rho} > 0, \qquad (8)$$

where in the region of $C$: $\left|\frac{\partial \mathbf{M}_-}{\partial \rho}\right| \approx \left|\frac{\partial \mathbf{M}_+}{\partial \rho}\right|$.

We note that (8) results in a steepness or skew effect (Hagan et al., 2002) which is further described in figure 5a. Furthermore, (8) implies that the surfaces of $\mathbf{M}_-$ and $\mathbf{M}_+$ intersect the zero plane at lines of intersection $\mathbf{A}_-$ and $\mathbf{A}_+$, as can be seen in figure 4. Due to the presence of a unique solution $C$, the lines $\mathbf{A}_-$ and $\mathbf{A}_+$ themselves intersect uniquely at $C$.

**Property 2.** *In the region of $C$:*

$$\frac{\partial (\mathbf{M}_- + \mathbf{M}_+)}{\partial \alpha} > 0, \qquad (9)$$

where $\frac{\partial \mathbf{M}_-}{\partial \alpha}$ and $\frac{\partial \mathbf{M}_+}{\partial \alpha}$ are monotone.

We note that (9) results in a curvature or smile effect (Hagan et al., 2002) which is further described in figure 5b. Critically, we state that (9) holds only in the region of $C$, however as stated within our subsequent discussion we assume that we are always able to observe the region of $C$ based on our initial discretisation of $\alpha$ (within $\bar{\alpha}$).

*2.4. Algorithm descriptions*

Our proposed calibration method operates by iteratively refining a two dimensional grid (or mesh) of $\alpha$ and $\rho$ values until the values $\alpha^*, \rho^*$ are obtained, an example of which is shown figure 6. Each iteration involves the construction of the matrices $\mathbf{M}_-$ and $\mathbf{M}_+$, as shown in (4) to (6). Parallelism is achieved during the construction of $\mathbf{M}_-$ and $\mathbf{M}_+$, whereby each element within such matrices is computed independently by multiple parallel threads.

In order to refine the grid of $\alpha$ and $\rho$ values for a subsequent iteration we require non-trivial information relating to the location of the solution $C$. This information is obtained through four principle steps, whereby the first step obtains lines of intersection $\mathbf{A}_-$ and $\mathbf{A}_+$. The second and third steps respectively obtain lower and upper bounds on $\alpha$ within $\bar{\alpha}$ and lower and upper bounds on $\rho$ within $\bar{\rho}$. The stated lower and upper bounds are designed to guarantee the bracketing of the solution $C$. Finally, the fourth step (which can be used optionally) involves estimating the precise location of the solution $C$. This enables the grid for a subsequent iteration to incorporate finer meshing or local mesh refinement around the estimate of $C$ (see figures 6b to 6f), hence improving the rate of convergence to the solution $C$. We now present a detailed description of the stated four steps.

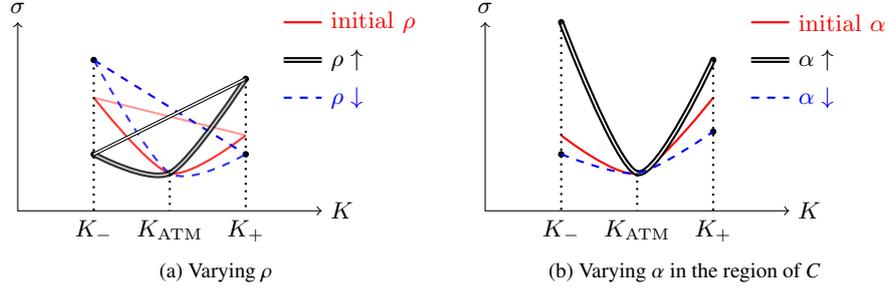

(a) Varying $\rho$

(b) Varying $\alpha$ in the region of $C$

Figure 5: (a) shows the SABR 'steepness' effect based on changes to $\rho$. For increasing/decreasing values of $\rho$, the gradient $\sigma(K_+) - \sigma(K_-)$ increases/decreases, this is achieved by decreasing/increasing $\sigma(K_-)$ and simultaneously increasing/decreasing $\sigma(K_+)$. (b) shows the SABR 'curvature' effect based on changes to $\alpha$ in the region of $C$. For higher/lower values of $\alpha$, the 'curvature' $\sigma(K_+) + \sigma(K_-)$ increases/decreases.

### 2.4.1. Step 1: Obtaining lines of intersection $\mathbf{A}_-$ and $\mathbf{A}_+$ (algorithm 1)

The lines $\mathbf{A}_-, \mathbf{A}_+$ represent lines of intersection between the surfaces of $\mathbf{M}_-, \mathbf{M}_+$ and the zero plane, examples of which are shown in figure 4. Within our calibration method we choose $\mathbf{A}_{(\cdot)}$ to be a vector of size $m$, where $m$ is equal to the size of $\bar{\alpha}$. Each value within $\bar{\alpha}$, referred to as $\alpha_i$, has a corresponding value $\mathbf{A}_{(\cdot)}(\alpha_i)$. The value $\mathbf{A}_-(\alpha_i)$ represents a column index of $\rho$ (within $\bar{\rho}$) such that $\mathbf{M}_-^{\alpha_i, \mathbf{A}_-(\alpha_i)} > 0$ and $\mathbf{M}_-^{\alpha_i, \mathbf{A}_-(\alpha_i)+1} < 0$ and the value $\mathbf{A}_+(\alpha_i)$ represents a column index of $\rho$ (within $\bar{\rho}$) such that $\mathbf{M}_+^{\alpha_i, \mathbf{A}_+(\alpha_i)} > 0$ and $\mathbf{M}_+^{\alpha_i, \mathbf{A}_+(\alpha_i)-1} < 0$.

We note that the lines $\mathbf{A}_-, \mathbf{A}_+$ represent intersections between the surfaces of $\mathbf{M}_-, \mathbf{M}_+$ and the zero plane based on an index of $\rho$ (as opposed to an index of $\alpha$), this is due to the unidirectional or monotone behaviour of $\frac{\partial \mathbf{M}_{(\cdot)}}{\partial \rho}$ which ensures intersection with the zero plane always occurs for a value of $\rho$. Due to (8) we expect the positive region of $\mathbf{M}_-$ to be for lower indices of $\rho$ (within $\bar{\rho}$), hence we initialise each value in $\mathbf{A}_-$ to represent the lowest index of $\rho$. Correspondingly we initialise each value in $\mathbf{A}_+$ to represent the highest index of $\rho$. A listing of this step is shown in algorithm 1.

**Algorithm 1:** Calculate $\mathbf{A}_-, \mathbf{A}_+$

1 Loop over all $\alpha$ values;
2 **for** $i = \alpha_1$ to $\alpha_m$ **do**
3     Initialise values;
4     $\mathbf{A}_-(i) \leftarrow \rho_1$;
5     $\mathbf{A}_+(i) \leftarrow \rho_n$;
6     Loop from low to high $\rho$ values;
7     **for** $j = \rho_1$ to $\rho_n$ **do**
8         **if** $\mathbf{M}_-^{i,j} > 0$ **then**
9             $\mathbf{A}_-(i) \leftarrow j$;
10     Loop from high to low $\rho$ values;
11     **for** $j = \rho_n$ to $\rho_1$ **do**
12         **if** $\mathbf{M}_+^{i,j} > 0$ **then**
13             $\mathbf{A}_+(i) \leftarrow j$;

### 2.4.2. Step 2: Calculating bounds for $\alpha$ (algorithm 2)

The second step brackets $\alpha^*$ within lower and upper $\alpha$ bounds $[\alpha_s, \alpha_f]$, where $\alpha_s, \alpha_f$ are $\alpha$ values within $\bar{\alpha}$.

In the region of $C$, due to (9) and since $\mathbf{M}_-$ and $\mathbf{M}_+$ are independent it follows that:

$$\frac{\partial \mathbf{M}_-}{\partial \alpha} + \frac{\partial \mathbf{M}_+}{\partial \alpha} > 0. \quad (10)$$

As a result of (10) three cases arise in the region of $C$:

1.
$$\frac{\partial \mathbf{M}_-}{\partial \alpha} > 0 \text{ and } \frac{\partial \mathbf{M}_+}{\partial \alpha} > 0. \quad (11)$$

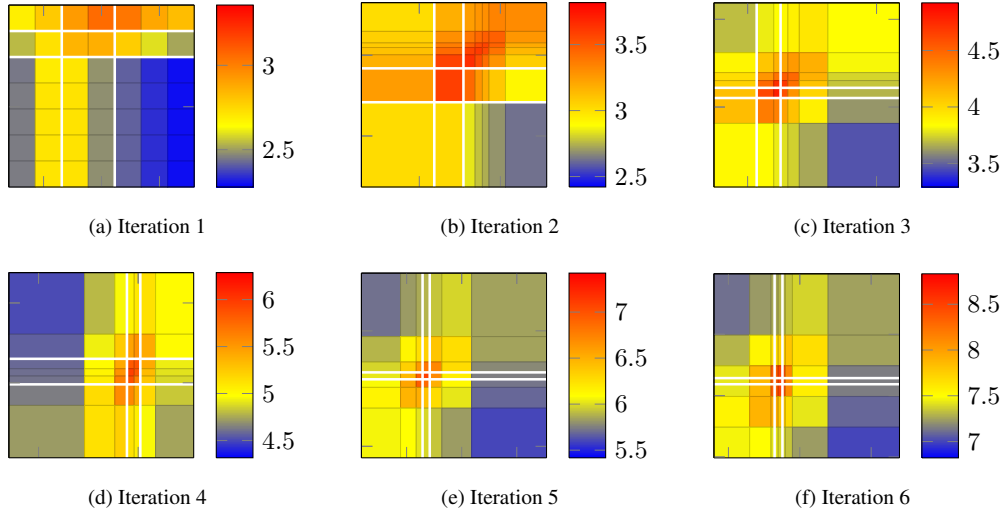

Figure 6: An example of the parallel SABR calibration method iterating to a solution. The *y* axis represents values of $\alpha$ and the *x* axis represents values of $\rho$. The colourbar value of each subgrid represents the error associated with the subgrid's top-left corner, where error is calculated as the RHS of (7). Error is reported as $-\log_{10}$ error (e.g. 8.5 represents an error of $10^{-8.5}$). The clear white lines within each grid represent bounds used for the next iteration. Local mesh refinement (algorithm 5) is not enabled during the first iteration.

**Algorithm 2:** Calculate $\alpha$ bounds $\alpha_s$ and $\alpha_f$

1. Initialise values;
2. $\alpha_s \leftarrow \alpha_1$;
3. $\alpha_f \leftarrow \alpha_m$;
4. Loop from low to high $\alpha$ values;
5. **for** $i = \alpha_1$ to $\alpha_m$ **do**
6.     **if** $\mathbf{A}_-(i) < \mathbf{A}_+(i) - 1$ **then**
7.        $\alpha_s \leftarrow i$;
8.     **if** $\mathbf{A}_-(i) < \mathbf{A}_+(i)$ **then**
9.        $\alpha_f \leftarrow i + 1$;
10. Ensure answer is not out of bounds;
11. $\alpha_f \leftarrow \mathrm{Min}(\alpha_f, \alpha_m)$;

Under this case when $\rho = \rho^*$ both $\mathbf{M}_-^{\alpha,\rho}$ and $\mathbf{M}_+^{\alpha,\rho}$ are negative when $\alpha < \alpha^*$ (and positive when $\alpha > \alpha^*$).

2. 
$$\frac{\partial \mathbf{M}_-}{\partial \alpha} < 0 \text{ and } \frac{\partial \mathbf{M}_+}{\partial \alpha} > \left|\frac{\partial \mathbf{M}_-}{\partial \alpha}\right|. \quad (12)$$

Under this case when $\rho = \rho^*$ only $\mathbf{M}_+^{\alpha,\rho}$ is negative when $\alpha < \alpha^*$ (and positive when $\alpha > \alpha^*$).

3. 
$$\frac{\partial \mathbf{M}_+}{\partial \alpha} < 0 \text{ and } \frac{\partial \mathbf{M}_-}{\partial \alpha} > \left|\frac{\partial \mathbf{M}_+}{\partial \alpha}\right|. \quad (13)$$

Under this case when $\rho = \rho^*$ only $\mathbf{M}_-^{\alpha,\rho}$ is negative when $\alpha < \alpha^*$ (and positive when $\alpha > \alpha^*$).

Furthermore we observe that *it is only when $\alpha < \alpha^*$ that both $\mathbf{M}_-^{\alpha,\rho}$ and $\mathbf{M}_+^{\alpha,\rho}$ can be negative, and only when $\alpha > \alpha^*$ that both $\mathbf{M}_-^{\alpha,\rho}$ and $\mathbf{M}_+^{\alpha,\rho}$ can be positive.*

The preceding statement is self-evident in case 1 and is demonstrated by figures 7a and 7b. In case 2, when $\rho = \rho^*$ and $\alpha < \alpha^*$, $\mathbf{M}_-^{\alpha,\rho}$ is positive, however due to (12) the positive magnitude of $\mathbf{M}_-^{\alpha,\rho}$ is dominated by the negative magnitude of $\mathbf{M}_+^{\alpha,\rho}$. Further, due to $\left|\frac{\partial \mathbf{M}_-}{\partial \rho}\right| \approx \left|\frac{\partial \mathbf{M}_+}{\partial \rho}\right|$ (property 1) and (8) we observe that a small increase in $\rho$ will shift $\mathbf{M}_-^{\alpha,\rho}$ into a negative region *before* shifting $\mathbf{M}_+^{\alpha,\rho}$ into a positive region, hence ensuring that both $\mathbf{M}_-^{\alpha,\rho}$ and $\mathbf{M}_+^{\alpha,\rho}$ are negative, this is demonstrated in figure 7c. A similar principle can be applied when $\alpha > \alpha^*$ and to case 3 (see figure 7d).

Our $\alpha$ bounds can now be determined as follows:

*A lower bound on $\alpha$, $\alpha_s$, is determined by the highest index of $\alpha_i$ (within $\bar{\alpha}$) such that $\mathbf{A}_-(\alpha_i) < \mathbf{A}_+(\alpha_i) - 1$. An upper bound on $\alpha$, $\alpha_f$, is determined by the lowest index of $\alpha_i$ (within $\bar{\alpha}$) such that $\mathbf{A}_-(\alpha_i) \geq \mathbf{A}_+(\alpha_i)$.*

To demonstrate the preceding statement, since $\alpha_s$ satisfies $\mathbf{A}_-(\alpha_s) < \mathbf{A}_+(\alpha_s) - 1$, within the rows $\mathbf{M}_-^{\alpha_s,(\cdot)}$ and $\mathbf{M}_+^{\alpha_s,(\cdot)}$ there are corresponding elements (i.e. elements with the same $\alpha, \rho$ index) that are both negative. Considering the corresponding elements $\mathbf{M}_-^{\alpha_s,\rho}$ and $\mathbf{M}_+^{\alpha_s,\rho}$ are negative, due to (8), any movement in $\rho$ will result in at least one of $\mathbf{M}_-^{\alpha_s,\rho}$ and $\mathbf{M}_+^{\alpha_s,\rho}$ becoming more negative, hence the solution $C$ cannot be obtained when $\alpha = \alpha_s$. This effect is shown in figure 8a. Recall, $\alpha_s$ is determined by the *highest* index of $\alpha_i$, this is because when $\alpha$ increases from being below $\alpha^*$ to greater than $\alpha^*$, both $\mathbf{M}_-^{\alpha,\rho}$ and $\mathbf{M}_+^{\alpha,\rho}$ go from being negative to positive, as was demonstrated earlier.

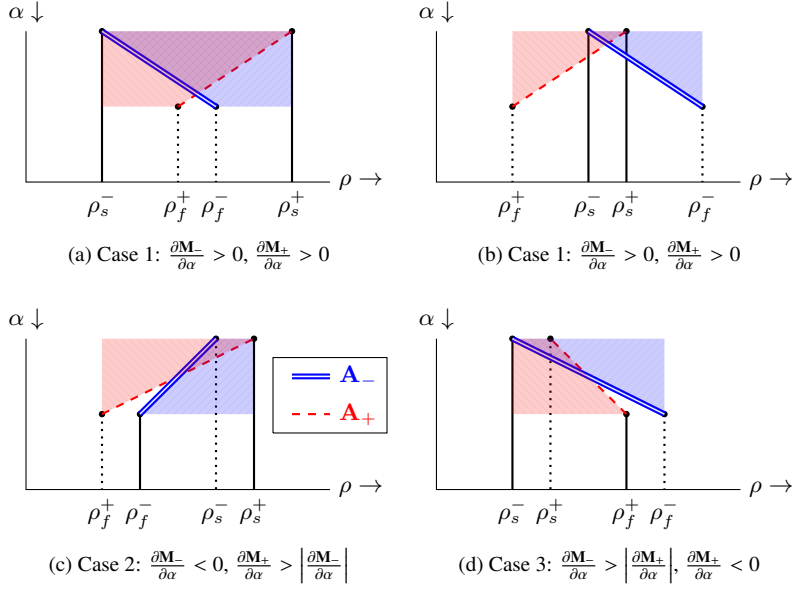

Figure 7: The shaded area bounded by $\mathbf{A}_-$ represents the region where $\mathbf{M}_-$ is negative, the shaded area bounded by $\mathbf{A}_+$ represents the region where $\mathbf{M}_+$ is negative. Cases 1 to 3 represent equations (11) to (13). Depicted $\alpha$ values are assumed to be in the region of $C$. Bounds for $\rho$ are given by the solid lines meeting the $x$ axis and have been calculated by the zero line (ZL) method (algorithm 3).

Also, since $\alpha_f$ satisfies $\mathbf{A}_-(\alpha_f) \geq \mathbf{A}_+(\alpha_f)$, within the rows $\mathbf{M}_-^{\alpha_f,(\cdot)}$ and $\mathbf{M}_+^{\alpha_f,(\cdot)}$ there are corresponding elements that are both positive. Considering the corresponding elements $\mathbf{M}_-^{\alpha_f,\rho}$ and $\mathbf{M}_+^{\alpha_f,\rho}$ are positive, due to (8), any movement in $\rho$ will result in at least one of $\mathbf{M}_-^{\alpha_f,\rho}$ and $\mathbf{M}_+^{\alpha_f,\rho}$ becoming more positive, hence the solution $C$ cannot be obtained when $\alpha = \alpha_f$. This effect is shown in figure 8c. Recall, $\alpha_f$ is determined by the *lowest* index of $\alpha_i$ as was described in the previous paragraph.

Rows represented by $\mathbf{M}_-^{\alpha,(\cdot)}$ and $\mathbf{M}_+^{\alpha,(\cdot)}$ such that $\alpha_s < \alpha < \alpha_f$ may contain the solution $C$ due to the presence of corresponding elements with opposite signs. Thus movements in $\rho$ will result in decreasing absolute error in both corresponding elements, an example of which is shown in figure 8b. Example $\alpha$ bounds are shown in figure 9a. A listing of this step is shown in algorithm 2.

#### 2.4.3. Step 3: Calculating bounds for $\rho$

The third step brackets $\rho^*$ within lower and upper $\rho$ bounds $[\rho_s, \rho_f]$, where $\rho_s, \rho_f$ are $\rho$ values within $\bar{\rho}$. Two methods exist for this purpose, firstly, the zero line (ZL) method, and secondly, the relative value (RV) method that is based on an additional property, shown in (16), relating to the region *very* close to $C$.

Bounds on $\rho$ are calculated by assuming that the previously calculated bounds on $\alpha$ capture the region of $C$ (in the $\alpha$ domain). In practice this is achieved by ensuring the $\alpha$ domain within $\bar{\alpha}$ is sufficiently discretised, whereby practical convergence examples are provided within section 2.5.

*The zero line (ZL) method (algorithm 3)*

Given $\mathbf{A}_{(\cdot)}$ and the $\alpha$ bounds $\alpha_s, \alpha_f$ we calculate $\rho$ indices $\rho_s^-, \rho_s^+, \rho_f^-, \rho_f^+$ where:

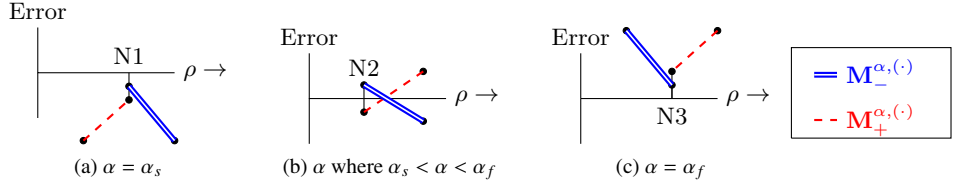

Figure 8: An illustration of $\alpha$ bounds $(\alpha_s, \alpha_f)$. The y axis represents an error value shown in (6). (a) shows the bound $\alpha_s$ where there is a node of $\rho$ (N1) such that both $\mathbf{M}_-^{\alpha_s,N1}$ and $\mathbf{M}_+^{\alpha_s,N1}$ have negative errors, where such negative errors are observed to decrease for changes in $\rho$. (c) shows the bound $\alpha_f$ where there is a node of $\rho$ (N3) such that both $\mathbf{M}_-^{\alpha_f,N3}$ and $\mathbf{M}_+^{\alpha_f,N3}$ have positive errors, where such positive errors are observed to increase for changes in $\rho$. (b) shows a value of $\alpha$ such that $\alpha_s < \alpha < \alpha_f$ where there is a node of $\rho$ (N2) such that $\mathbf{M}_-^{\alpha,N2}$ and $\mathbf{M}_+^{\alpha,N2}$ have errors of opposite signs, where such absolute errors are observed to decrease for changes $\rho$, consequently using this value of $\alpha$ it is thus possible for changes in $\rho$ to result in the solution $\mathbf{M}_-^{\alpha,\rho} = 0$ and $\mathbf{M}_+^{\alpha,\rho} = 0$.

$$\rho_s^- = \mathbf{A}_-(\alpha_s), \qquad \rho_f^- = \mathbf{A}_-(\alpha_f), \qquad (14)$$
$$\rho_s^+ = \mathbf{A}_+(\alpha_s), \qquad \rho_f^+ = \mathbf{A}_+(\alpha_f). \qquad (15)$$

Our $\rho$ bounds can now be determined as follows based on possible signs of $\frac{\partial \mathbf{M}_-}{\partial \alpha}$ and $\frac{\partial \mathbf{M}_\pm}{\partial \alpha}$ (as listed within (11) to (13)):

$\rho_s^-$ is a lower bound for $\rho$ when $\frac{\partial \mathbf{M}_-}{\partial \alpha} > 0$. $\rho_f^-$ is a lower bound for $\rho$ when $\frac{\partial \mathbf{M}_-}{\partial \alpha} < 0$. $\rho_s^+$ is an upper bound for $\rho$ when $\frac{\partial \mathbf{M}_+}{\partial \alpha} > 0$. $\rho_f^+$ is an upper bound for $\rho$ when $\frac{\partial \mathbf{M}_+}{\partial \alpha} < 0$.

To demonstrate the preceding statement we first attempt to show that when $\frac{\partial \mathbf{M}_-}{\partial \alpha} > 0$ the solution $C$ (where $\mathbf{M}_{(\cdot)}^{\alpha^*,\rho^*} = 0$) can only be obtained with values of $\rho$ that are greater than a lower bound $\rho_s^-$ (having already established that $\alpha_s$ is a lower bound on $\alpha$). As shown in section 2.4.1, $\mathbf{M}_-^{\alpha_s,\rho_s^-} > 0$ and $\mathbf{M}_-^{\alpha_s,\rho_s^-+1} < 0$. Due to (8) there exists a value of $\rho$ namely $\rho_0$ such that $\mathbf{M}_-^{\alpha_s,\rho_0} = 0$, where $\rho_0 > \rho_s^-$. Further, given $\frac{\partial \mathbf{M}_-}{\partial \alpha} > 0$, an increase in $\alpha_s$ of $\Delta\alpha$ (towards $\alpha^*$) results in $\mathbf{M}_-^{\alpha_s+\Delta\alpha,\rho_0} > 0$. Further due to (8), an increase in $\rho_0$ of $\Delta\rho$ results in $\mathbf{M}_-^{\alpha_s+\Delta\alpha,\rho_0+\Delta\rho} = 0$, where $\rho_0 + \Delta\rho > \rho_0 > \rho_s^-$. We have thus demonstrated that the solution $C$ (where $\mathbf{M}_-^{\alpha^*,\rho^*} = 0$) can only be obtained by values of $\rho$ greater than a lower bound $\rho_s^-$. We note how this derivation relies on $\frac{\partial \mathbf{M}_-}{\partial \alpha} > 0$ being monotone as stated in property 2. The remaining bounds $(\rho_s^+, \rho_f^-, \rho_f^+)$ are demonstrated by reapplying the above method.

A depiction of resultant bounds is shown in figure 7. For case 1, based on (11), lower and upper $\rho$ bounds are given by $\rho_s^-$ and $\rho_s^+$ respectively (see figures 7a and 7b). This occurs when $\rho_s^- < \rho_f^-$ and $\rho_s^+ > \rho_f^+$, where the lines $\mathbf{A}_-$ and $\mathbf{A}_+$ are shown to slope in opposite directions. For case 2, based on (12), the lower $\rho$ bound is given by $\rho_f^-$ rather than $\rho_s^-$ (see figure 7c). This occurs when $\rho_s^- > \rho_f^-$ where the lines $\mathbf{A}_-$ and $\mathbf{A}_+$ slope down in the same leftward direction. For case 3, based on (13), the upper $\rho$ bound is given by $\rho_f^+$ rather than $\rho_s^+$ (see figure 7d). This occurs when $\rho_s^+ < \rho_f^+$ where the lines $\mathbf{A}_-$ and $\mathbf{A}_+$ slope down in the same rightward direction. In instances when $\rho_s^- = \rho_f^-$ or $\rho_s^+ = \rho_f^+$, all cases produce identical lower or upper bounds respectively.

Example $\rho$ bounds generated by the zero line (ZL) method are shown in figures 9b, 9c and 9d. A listing of this step is shown in algorithm 3.

---

**Algorithm 3:** Calculate $\rho$ bounds $\rho_s$ and $\rho_f$ based on the zero line (ZL) method

1 $\rho_s^- \leftarrow \mathbf{A}_-(\alpha_s)$;
2 $\rho_s^+ \leftarrow \mathbf{A}_+(\alpha_s)$;
3 $\rho_f^- \leftarrow \mathbf{A}_-(\alpha_f)$;
4 $\rho_f^+ \leftarrow \mathbf{A}_+(\alpha_f)$;
5 Bounds as in figure 7c;
6 **if** $\rho_s^- \geq \rho_f^-$ **then**
7 $\quad \rho_s \leftarrow \rho_f^-$;
8 $\quad \rho_f \leftarrow \rho_s^+$;
9 Bounds as in figure 7d;
10 **else if** $\rho_s^+ \leq \rho_f^+$ **then**
11 $\quad \rho_s \leftarrow \rho_s^-$;
12 $\quad \rho_f \leftarrow \rho_f^+$;
13 Bounds as in figures 7a, 7b;
14 **else**
15 $\quad \rho_s \leftarrow \rho_s^-$;
16 $\quad \rho_f \leftarrow \rho_s^+$;

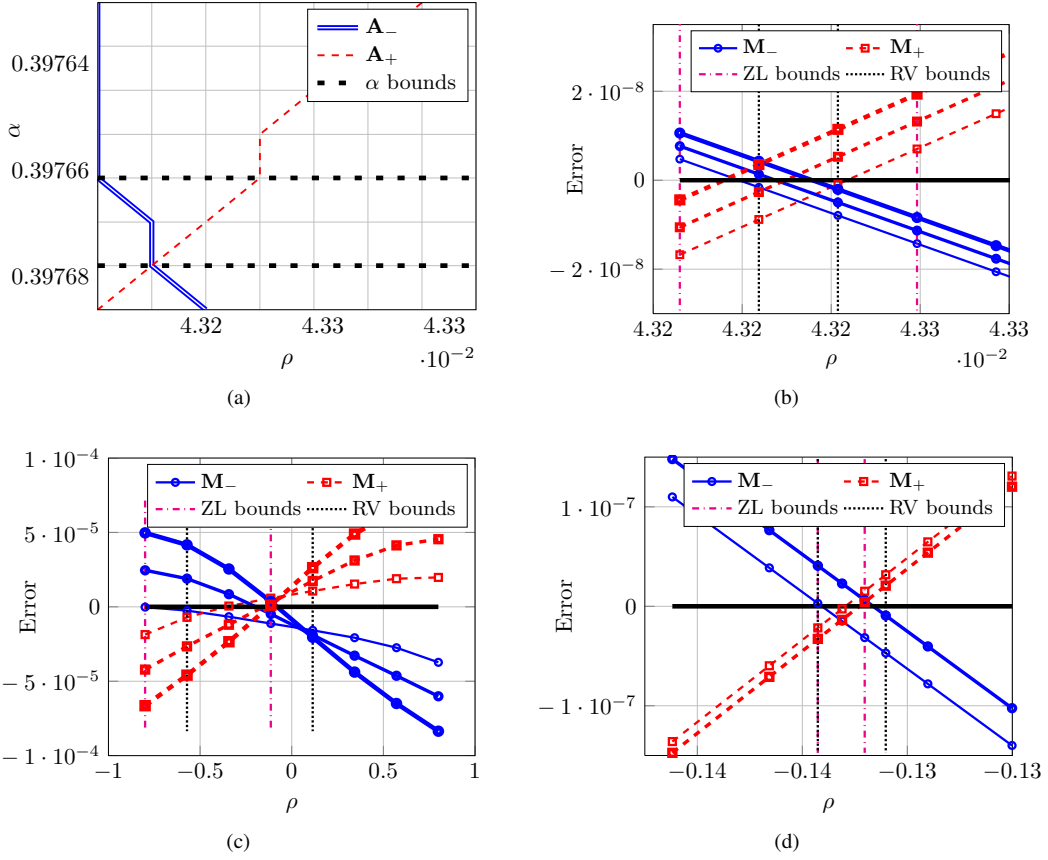

Figure 9: (a) shows the lines $\mathbf{A}_-$ and $\mathbf{A}_+$ alongside $\alpha$ bounds generated by algorithm 2. (b), (c) and (d) show error values within $\mathbf{M}_-$ and $\mathbf{M}_+$ calculated by (6) for changing values of $\rho$ within lower and upper bounds of $\alpha$. The faintest lines for $\mathbf{M}_-$ and $\mathbf{M}_+$ represent the lower $\alpha$ bound, $\alpha_s$, higher $\alpha$ values are represented by lines of increased thickness, whereby the thickest lines for $\mathbf{M}_-$ and $\mathbf{M}_+$ represent the upper $\alpha$ bound, $\alpha_f$. (a) and (b) correspond to identical examples of $\mathbf{M}_-$ and $\mathbf{M}_+$. (c) and (d) are based on the same data set as figure 4 but with different discretisation, where (c) shows an initial iteration enclosing a large area away from $C$ and (d) shows a subsequent iteration enclosing an area close to $C$.

*The relative value (RV) method (algorithm 4)*

Within this algorithm we make use of an additional property of our error matrices $\mathbf{M}_-$ and $\mathbf{M}_+$:

**Property 3.** *In the region very close to C:*

$$\frac{\partial^2 \mathbf{M}_{(\cdot)}}{\partial \alpha^2} \approx 0, \qquad \frac{\partial^2 \mathbf{M}_{(\cdot)}}{\partial \rho^2} \approx 0. \qquad (16)$$

Our previously described zero line (ZL) method (algorithm 3) was based on the vectors $\mathbf{A}_{(\cdot)}$ which contain information relating to the position of positive and negative elements within $\mathbf{M}_{(\cdot)}$ (through obtaining the line of intersection between the surfaces of $\mathbf{M}_{(\cdot)}$ and the zero plane). Further information can be obtained by evaluating differences between corresponding elements (i.e. elements with the same $\alpha, \rho$ index) within $\mathbf{M}_-$ and $\mathbf{M}_+$.

Using our $\alpha$ bounds $\alpha_s, \alpha_f$, we define $I_s$ as the maximum $\rho$ index (within $\bar{\rho}$) such that $\mathbf{M}_-^{\alpha_s, I_s} > \mathbf{M}_+^{\alpha_s, I_s}$ (where due to (8) $\mathbf{M}_-^{\alpha_s, I_s+1} < \mathbf{M}_+^{\alpha_s, I_s+1}$) and define $I_f$ as the maximum $\rho$ index (within $\bar{\rho}$) such that $\mathbf{M}_-^{\alpha_f, I_f} > \mathbf{M}_+^{\alpha_f, I_f}$ (where due to (8) $\mathbf{M}_-^{\alpha_f, I_f+1} < \mathbf{M}_+^{\alpha_f, I_f+1}$).

Our $\rho$ bounds can now be determined as follows:

*In the region very close to C, the $\rho$ values corresponding to* $\mathrm{Min}(I_s, I_f)$ *and* $\mathrm{Max}(I_s, I_f) + 1$ *are lower and upper bounds on $\rho$ respectively.*

To demonstrate the preceding statement we initially identify that in the region of $C$, due to (9), $\mathbf{M}_-^{\alpha_s,(\cdot)} + \mathbf{M}_+^{\alpha_s,(\cdot)} < 0$ and that $\mathbf{M}_-^{\alpha_s,I_s} > \mathbf{M}_+^{\alpha_s,I_s}$. Consequently due to (8) there exists a value of $\rho$, indexed by $I_s + \Delta\rho$, where $I_s < I_s + \Delta\rho < I_s + 1$, such that:

$$\mathbf{M}_-^{\alpha_s,I_s+\Delta\rho} = \mathbf{M}_+^{\alpha_s,I_s+\Delta\rho} < 0. \tag{17}$$

Similarly, there exists a value of $\rho$, namely $I_f + \Delta\rho$, where $I_f < I_f + \Delta\rho < I_f + 1$, such that:

$$\mathbf{M}_-^{\alpha_f,I_f+\Delta\rho} = \mathbf{M}_+^{\alpha_f,I_f+\Delta\rho} > 0. \tag{18}$$

Also, at the solution $C$:

$$\mathbf{M}_-^{\alpha^*,\rho^*} = \mathbf{M}_+^{\alpha^*,\rho^*} = 0. \tag{19}$$

The movement of error from $\mathbf{M}_{(\cdot)}^{\alpha_s,I_s+\Delta\rho} < 0$ to $\mathbf{M}_{(\cdot)}^{\alpha_f,I_f+\Delta\rho} > 0$ is based on changes to both $\alpha$ and $\rho$. Due to (16), it follows that the above traversal from $\mathbf{M}_{(\cdot)}^{\alpha_s,I_s+\Delta\rho} < 0$ to $\mathbf{M}_{(\cdot)}^{\alpha_f,I_f+\Delta\rho} > 0$ is based on a fixed unidirectional rate of change with respect to $\alpha$ and $\rho$, whereby the traversal includes the point $\mathbf{M}_{(\cdot)}^{\alpha^*,\rho^*}$. Therefore if $I_s + \Delta\rho < I_f + \Delta\rho$ it follows that $I_s + \Delta\rho < \rho^* < I_f + \Delta\rho$, which results in the $\rho$ bounds $[I_s, I_f + 1]$. Alternately, if $I_s + \Delta\rho > I_f + \Delta\rho$, $\rho$ bounds are given as $[I_f, I_s+1]$. Thus we set our lower bound $\rho_s$ as $\text{Min}(I_s, I_f)$ and upper bound $\rho_f$ as $\text{Max}(I_s, I_f) + 1$.

Example $\rho$ bounds generated by the relative value (RV) method are shown in figures 9b, 9c and 9d. A listing of this step is show in algorithm 4.

### 2.4.4. Step 4: Local mesh refinement (algorithm 5)

This step is used optionally and involves estimating the solution $C$ (i.e. $\alpha^*, \rho^*$) as $\alpha^{\text{Target}}, \rho^{\text{Target}}$. The objective is to build a locally refined mesh that is fine near the estimated solution, and progressively coarse away from the estimated solution, thereby attempting to improve the rate of convergence to $C$.

The location of $\alpha^{\text{Target}}, \rho^{\text{Target}}$ is based on the linearity expressed in (16). As mentioned in the discussion of the relative value (RV) method (algorithm 4), the traversal of error from (17) to (19) to (18) is based on a constant unidirectional rate of change with respect to $\alpha$ and $\rho$. As such we can extrapolate the likely position of $\alpha^{\text{Target}}, \rho^{\text{Target}}$. The method works as follows:

Firstly, we obtain a vector $I$ where each element $I_i$ corresponds to a value $\alpha_i$ within the range $\alpha_s, \cdots, \alpha_f$ and is populated with the maximum $\rho$ index (within $\bar\rho$) such that $\mathbf{M}_-^{\alpha_i,I_i} > \mathbf{M}_+^{\alpha_i,I_i}$ (where due to (8) $\mathbf{M}_-^{\alpha_i,I_i+1} < \mathbf{M}_+^{\alpha_i,I_i+1}$). We note that in the relative value (RV) method we obtained $I_s$ and $I_f$ which are values of $I$ corresponding to $\alpha_s$ and $\alpha_f$ respectively.

Secondly, we define the following six vectors with indices $i$ and which are of size equal to $I$:

$$\begin{aligned}\bar\epsilon_-^1(i) &= \mathbf{M}_-^{\alpha_i,I_i}, & \bar\epsilon_-^2(i) &= \mathbf{M}_-^{\alpha_i,I_i+1}, \\ \bar\epsilon_+^1(i) &= \mathbf{M}_+^{\alpha_i,I_i}, & \bar\epsilon_+^2(i) &= \mathbf{M}_+^{\alpha_i,I_i+1}, \\ \bar\rho_I(i) &= \bar\rho(I_i), & \bar\rho_{I+1}(i) &= \bar\rho(I_i+1).\end{aligned} \tag{20}$$

We further define the vectors $\rho_I$ and $\epsilon_I$ (note the absence of a horizontal line) which are also of size equal to $I$, where $\rho_I$ represents *estimated* values of $\rho$ at which $\mathbf{M}_-^{\alpha,\rho} = \mathbf{M}_+^{\alpha,\rho}$ and $\epsilon_I$ represents the corresponding *estimated* error values. Each estimate within the vectors $\rho_I$ and $\epsilon_I$ is representative of an $\alpha$ value within $\alpha_s, \cdots, \alpha_f$. An example of the pairs $(\rho_I, \epsilon_I)$ is shown in figure 10. $\rho_I$ and $\epsilon_I$ are obtained by using the intermediate vectors $a_-, a_+, b_-, b_+$ (also of size equal to $I$) as per the following element-by-element arithmetic:

---

**Algorithm 4:** Calculate $\rho$ bounds $\rho_s$ and $\rho_f$ based on the relative value (RV) method

1. Initialise values;
2. $I_s \leftarrow \rho_1$;
3. $I_f \leftarrow \rho_n$;
4. Loop from low to high $\rho$ values;
5. **for** $i = \rho_1$ *to* $\rho_n$ **do**
6.     **if** $\mathbf{M}_-^{\alpha_s,i} > \mathbf{M}_+^{\alpha_s,i}$ **then**
7.         $I_s \leftarrow i$;
8.     **if** $\mathbf{M}_-^{\alpha_f,i} > \mathbf{M}_+^{\alpha_f,i}$ **then**
9.         $I_f \leftarrow i$;
10. $\rho_s \leftarrow \text{Min}(I_s, I_f)$;
11. $\rho_f \leftarrow \text{Max}(I_s, I_f) + 1$;
12. Ensure answer is not out of bounds;
13. $\rho_f \leftarrow \text{Min}(\rho_f, \rho_n)$;

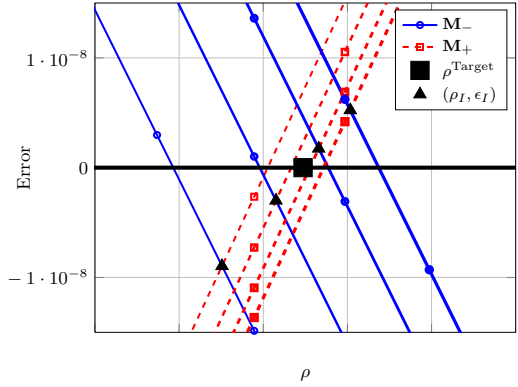

Figure 10: Depiction of the local mesh refinement step (algorithm 5) obtaining a guess for $\rho^{\text{Target}}$ by estimating $\rho_I, \epsilon_I$, where $\rho_I, \epsilon_I$ are points that represent where $\mathbf{M}_-^{\alpha,\rho} = \mathbf{M}_+^{\alpha,\rho}$. The intersection of the line connecting elements of $\rho_I, \epsilon_I$ with the $x$ axis is taken as a guess for $\rho^{\text{Target}}$. The $y$ axis represents an error value shown in (6).

$$a_-(i) = \frac{\bar{\epsilon}_-^1(i) - \bar{\epsilon}_-^2(i)}{\bar{\rho}_I(i) - \bar{\rho}_{I+1}(i)}, \quad b_-(i) = \bar{\epsilon}_-^1(i) - a_-(i) \times \bar{\rho}_I(i),$$
$$a_+(i) = \frac{\bar{\epsilon}_+^1(i) - \bar{\epsilon}_+^2(i)}{\bar{\rho}_I(i) - \bar{\rho}_{I+1}(i)}, \quad b_+(i) = \bar{\epsilon}_+^1(i) - a_+(i) \times \bar{\rho}_I(i),$$
$$\rho_I(i) = -\frac{b_+(i) - b_-(i)}{a_+(i) - a_-(i)}, \quad \epsilon_I(i) = a_-(i) \times \rho_I(i) + b_-(i),$$
$$= a_+(i) \times \rho_I(i) + b_+(i). \quad (21)$$

Next, we obtain the index $i^*$ that corresponds to the minimum error within $\epsilon_I$, alongside the index $i^{**}$ that corresponds to the neighbour of $i^*$ that satisfies:

$$\text{sign}(\epsilon_I(i^*)) \neq \text{sign}(\epsilon_I(i^{**})). \quad (22)$$

Finally, we are able to identify $\rho^{\text{Target}}$ (seen in figure 10) and $\alpha^{\text{Target}}$ as follows using the intermediate variables $a_\rho, b_\rho, a_\alpha, b_\alpha$, where $i_1 = \text{Min}(i^*, i^{**})$ and $i_2 = \text{Max}(i^*, i^{**}) = i_1 + 1$:

$$a_\rho = \frac{\epsilon_I(i_1) - \epsilon_I(i_2)}{\rho_I(i_1) - \rho_I(i_2)}, \quad b_\rho = \epsilon_I(i_1) - a_\rho \times \rho_I(i_1),$$
$$a_\alpha = \frac{\epsilon_I(i_1) - \epsilon_I(i_2)}{\bar{\alpha}(i_1) - \bar{\alpha}(i_2)}, \quad b_\alpha = \epsilon_I(i_1) - a_\alpha \times \bar{\alpha}(i_1),$$
$$\rho^{\text{Target}} = -\frac{b_\rho}{a_\rho}, \quad \alpha^{\text{Target}} = -\frac{b_\alpha}{a_\alpha}. \quad (23)$$

Consequently we are able to build a two dimensional locally refined mesh that is centred at $\alpha^{\text{Target}}, \rho^{\text{Target}}$ and bounded by $(\alpha_s, \alpha_f)$ and $(\rho_s, \rho_f)$ as shown in figures 6b to 6f.

| Accuracy | 1e-6 | | 1e-8 | | 1e-10 | |
|---|---|---|---|---|---|---|
| | $t$ | Iter | $t$ | Iter | $t$ | Iter |
| GPU | 5.1 | 4.9 | 6.7 | 6.8 | 7.9 | 8.0 |
| CPU | 17.0 | 15.5 | 24.1 | 20.5 | 42.8 | 38.5 |
| Speedup | 3.3 | | 3.6 | | 5.4 | |

Table 1: Performance of the parallel SABR calibration method implemented on a GPU and a typical gradient descent algorithm implemented on a CPU for a problem size of 192 calibrations. $t$ = total execution time (ms), Iter = iterations. GPU iterations represent the average number of successive grid iterations. CPU iterations represent the average combined number of gradient updates on $\alpha$ and $\rho$. GPU results are based on the relative value (RV) method with local mesh refinement (V3). Speedup reported as CPU time / GPU time. Accuracy is calculated as the RHS of (7). CPU results are based on a single threaded implementation.

### 2.5. Performance results

Results are generated using the system properties shown in Appendix A. Our parallel calibration method is implemented on GPUs whereby each individual calibration is conducted by a different GPU 'thread block'. A thread block is viewed as a collection of parallel threads where the programmer is able to make use of a shared memory space and set synchronisation points. The use of a separate GPU thread block for each individual calibration is based on our chosen category of derivatives, which comprises of multiple coupons resulting in upto several hundred individual calibrations for pricing a single derivative.

Within our implementation the size of each GPU thread block is chosen to equal the number of elements within $\mathbf{M}_{(\cdot)}$ shown in (5), whereby each element within $\mathbf{M}_{(\cdot)}$ is calculated by a separate parallel thread. After the creation of $\mathbf{M}_{(\cdot)}$ all threads are synchronised and a single thread undertakes the sequential steps 1 to 4 listed in section 2.4. We note that the creation of $\mathbf{M}_{(\cdot)}$ vastly dominates the execution time of steps 1 to 4, thus minimising the performance penalty of using a single thread.

Our results depict the following changing parameters:

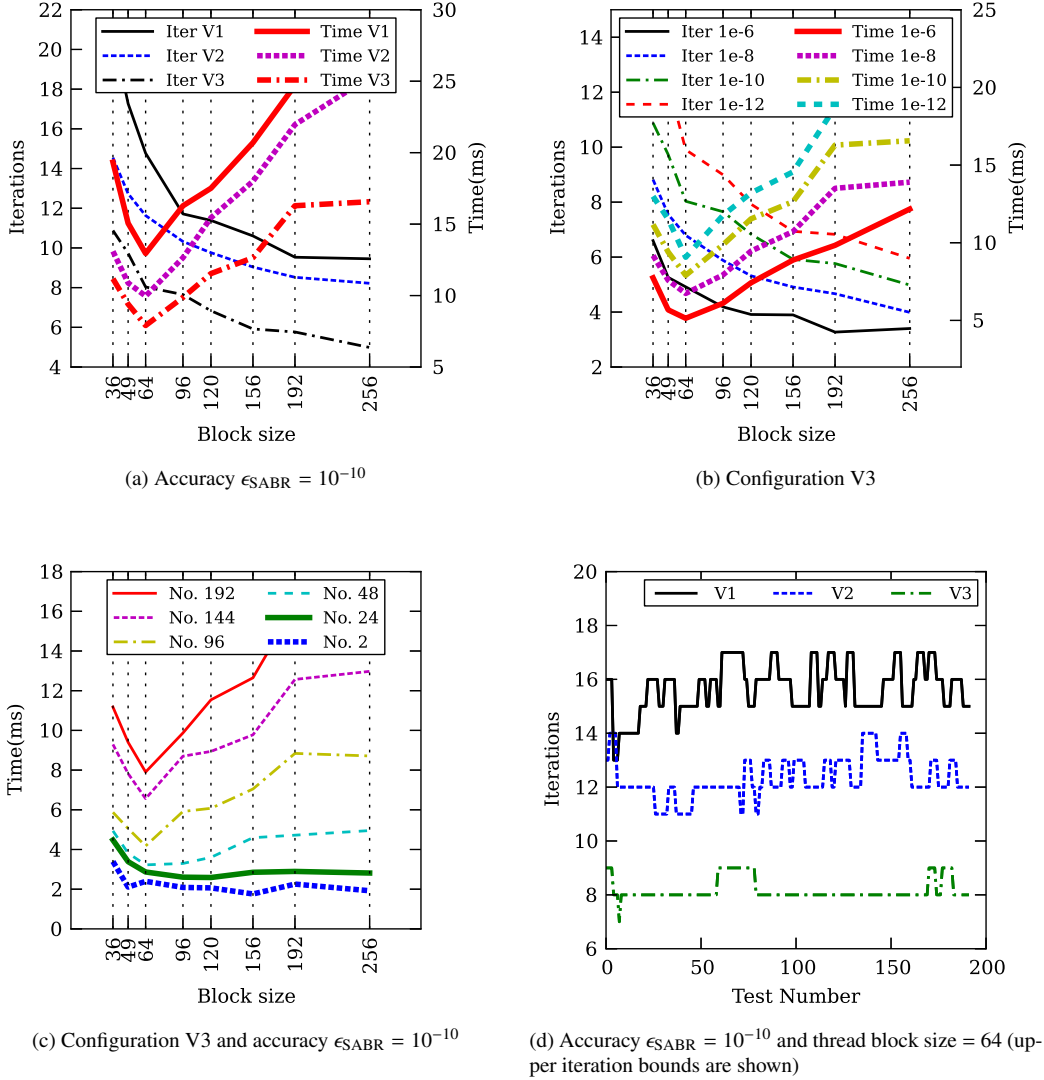

Figure 11: Performance of the parallel SABR calibration method for a typical problem size of 192 calibrations (except (c)). Iter = average iterations, Time = total kernel execution time (ms), No. = the number of calibrations conducted. V1/2/3 = configuration based on 1. zero line (ZL) method, 2. relative value (RV) method, 3. RV method with local mesh refinement. Block size = GPU thread block size. Accuracy 1e-x = $\epsilon_{SABR}$ as calculated by the RHS of (7).

Firstly, our GPU thread block size varies from 36 to 256 threads. Thread blocks are arranged with *x* and *y* coordinates representing a grid of $\alpha$ and $\rho$ values as shown in (5). Thread blocks are arranged to have (as far as possible) an equal number of threads within the *x* and *y* coordinates, for example a thread block with 64 threads represents a two dimensional grid of 8 × 8 threads.

Secondly, we vary the number of individual calibrations (or equivalently the number of GPU thread blocks launched) from 2 to 192. Higher numbers of individual calibrations did not change the algorithm dynamics (specifically the optimum thread block size as shown in figure 11c).

Thirdly, the calibration accuracy level $\epsilon_{SABR}$ calculated as the RHS of (7) varies from $10^{-6}$ to $10^{-12}$.

Finally, we vary our parallel calibration method to op-erate under different configurations relating to steps 3 and 4 listed in section 2.4. In all cases however, the first iteration undertakes step 3 using the zero line (ZL) method and does not undertake step 4. This is because the relative value (RV) method of step 3 and local mesh refinement of step 4 rely on being very close to the region of *C* which is assumed to be unavailable during the first iteration. For subsequent iterations the following configurations are used:

**V1** represents continued usage of the zero line (ZL) method in step 3 (step 4 is not undertaken).

**V2** represents usage of the relative value (RV) method in step 3 (step 4 is not undertaken).

**V3** represents usage of the relative value (RV) method in step 3 **and** local mesh refinement in step 4.

In terms of performance the zero line (ZL) method was always inferior to the relative value (RV) method, hence only the relative value (RV) method was used in conjunction with local mesh refinement. GPU thread blocks of size smaller than 36 did not converge for all of our test cases and have thus not been included within our analysis. Non-convergence was due to an insufficient initial discretisation of $\bar{\alpha}$ which led to the region of *C* not being captured in the initial iteration.

In figure 11a we show results when $\epsilon_{\text{SABR}} = 10^{-10}$ and observe that V3 using 64 threads has the best performance. Figure 11b shows results from V3 and depicts the effect on execution time and iterations needed when using different values of $\epsilon_{\text{SABR}}$. In figure 11c we vary the number of calibrations scheduled on the GPU from 2 to 192. We observe that when the number of calibrations scheduled is very low, the dominant factor in execution time is the number of required iterations whereby *larger thread blocks reduce the number of required iterations which consequently reduces execution time*. As the number of calibrations scheduled increase however, the optimum execution time is a compromise between the number of required iterations *and* the number of concurrent thread blocks scheduled within the GPU processor whereby *smaller thread blocks result in more concurrent thread blocks scheduled on the GPU which consequently reduces execution time*. The use of 64 threads within a thread block is shown to offer the best overall performance. Figure 11d shows results when $\epsilon_{\text{SABR}} = 10^{-10}$ and provides an overview of the number of iterations needed for each calibration configuration where V3 is shown to have the best performance.

Our GPU implementation is also compared briefly against a single threaded CPU implementation that performs SABR calibration based on a typical three dimensional gradient descent algorithm, results of which are shown in table 1. Results indicate that the use of the relative value (RV) method with local mesh refinement (V3) can outperform a single threaded CPU implementation by a factor of 6 to 8 ×. Additionally, our GPU implementation is shown to offer better performance for higher accuracies of $\epsilon_{\text{SABR}}$. This is attributed to a better final rate of convergence within our chosen V3 configuration, an example of which is shown in figure 6. As stated, within V3 an initial iteration uses the zero line (ZL) method, subsequently the relative value (RV) method with local mesh refinement is employed. Since local mesh refinement is designed for the region very close to *C*, several iterations are needed before $\alpha$ and $\rho$ bounds are based on the finest subgrid (as seen in figures 6e and 6f), at this point subsequent convergence settles to a linear rate dependent on the domain size of the finest subgrid. The example in figure 6 suggests a final linear convergence rate of $O\left(10^{-1.5}\right)$.

## 3. Cumulative probability lookup tables

This section will present parallel algorithms to form lookup tables consisting of cumulative probability distributions. A typical cumulative probability distribution is shown in figure 12a.

### 3.1. Background

Given a random variable, namely a future asset price or value $F(T)$, $P(x)$ is chosen to represent a cumulative probability which is defined as the probability that $F(T)$ will be assigned a value less than or equal to a particular asset value *x* such that:

$$P(x) = \text{Prob}(F(T) \leq x). \quad (24)$$

Within our subsequent algorithms the future asset price $F(T)$ is assumed to have a probability distribution implied by the SABR model, presented in section 2, however our algorithms are applicable to all probability distributions that exhibit clustering effects as described later in this section.

As part of our derivative pricing model we undertake a large Monte Carlo simulation, within which we are given cumulative probabilities $P(x)$ and are required to

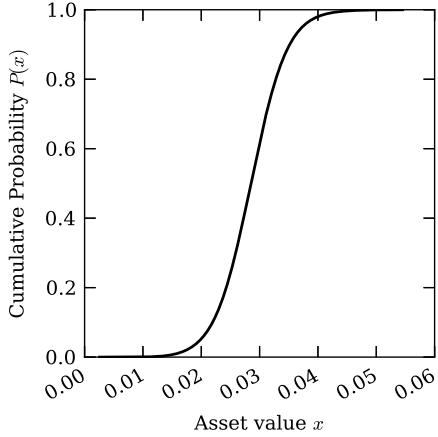
(a) Example cumulative probability distribution

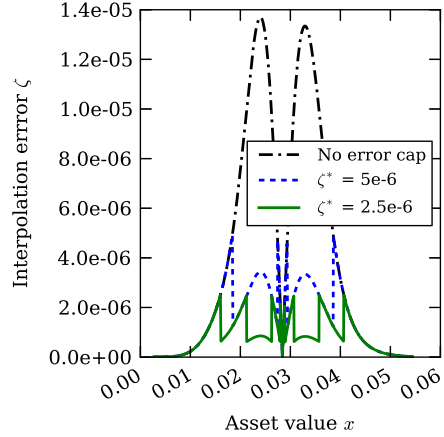
(b) Example linear interpolation error with error caps $\zeta^*$

Figure 12: (a) shows an example cumulative probability distribution. (b) shows the corresponding linear interpolation error based on (27). For the no error cap case the maximum interpolation error is centred around the greatest curvature within (a). We further observe that the maximum interpolation error is always clustered around a series of peaks.

calculate the corresponding asset value $x$. That is we are required to calculate the inverse probability:

$$x = P^{-1}(P(x)), \quad (25)$$

where $P^{-1}(P(\cdot)) = (\cdot)$. Using the SABR model it is possible to calculate $P(\cdot)$ cheaply analytically, however calculating an inverse $P^{-1}(\cdot)$ requires a computationally expensive iterative method as no analytic inverse exists. Within our stated Monte Carlo simulation we are required to conduct numerous unique evaluations of $P^{-1}(\cdot)$ equivalent to the size of the Monte Carlo simulation $s$, where $s$ is typically of $O(10^5$ to $10^8)$. Since the relationship $x \xrightarrow{P} P(x)$ (and conversely $P(x) \xrightarrow{P^{-1}} x$) is one-to-one and monotone it is typical to reduce computational effort by approximating (25) through a lookup table of size $N$, where $N < s$. Such a lookup table consists of an array $\bar{x}$ containing $N$ discrete ordered values of $x$ and an array $\bar{P}_x$ containing corresponding values of $P(x)$. To calculate an inverse as shown in (25) we use the following linear interpolation approach: firstly, the closest entries to our input $P(x)$ within the array $\bar{P}_x$ are found, namely $\bar{P}_x(a)$ and $\bar{P}_x(b)$ (e.g. through a binary search). Subsequently we interpolate between corresponding values within the array $\bar{x}$, namely $\bar{x}(a)$ and $\bar{x}(b)$, such that (25) is approximated as (where $a < b$):

$$x = \bar{x}(a) + \frac{\bar{x}(b) - \bar{x}(a)}{\bar{P}_x(b) - \bar{P}_x(a)}(P(x) - \bar{P}_x(a)). \quad (26)$$

3.2. Interpolation error

Lookup table linear interpolation error $\zeta$ can be practically estimated by the absolute difference between an interpolated mid-point value and an analytic mid-point value. Assuming the size of $\bar{x}$ and $\bar{P}_x$ is $N$, an array consisting of such mid-point interpolation errors $\bar{\zeta}$ will be of size $N-1$. Each element of this array can be shown as:

$$\bar{\zeta}_i = \left| P\left(\frac{\bar{x}(i) + \bar{x}(i+1)}{2}\right) - \frac{\bar{P}_x(i) + \bar{P}_x(i+1)}{2} \right|,$$
$$\text{where } i \in \{1, \cdots, N-1\}. \quad (27)$$

It is possible to cap the interpolation error calculated in (27) to a maximum value $\zeta^*$ (as shown in figure 12b). In doing so we refine the discretisation of $x$ within the array $\bar{x}$ (and correspondingly the array $\bar{P}_x$) when the interpolation error calculated in (27) is greater than $\zeta^*$, this is shown in figure 13. Consequently the lookup table size $N$ may vary for different values of $\zeta^*$. It is further noted (as shown within figure 12b) that maximum interpolation error is clustered around a series of peaks, critically, our subsequent algorithms are optimised for probability distributions that exhibit this phenomenon.

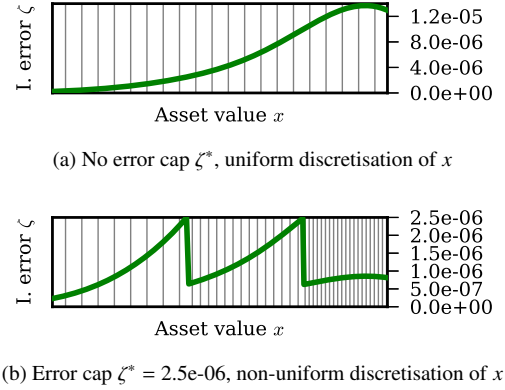

(a) No error cap $\zeta^*$, uniform discretisation of $x$

(b) Error cap $\zeta^* = 2.5\text{e-}06$, non-uniform discretisation of $x$

Figure 13: I. error = interpolation error based on (27). Plots are based on the same data set as figure 12b. In (b) refining the discretisation of $x$ results in lower interpolation error $\zeta$.

### 3.3. Algorithm descriptions

Within this section we present several algorithms used to create cumulative probability lookup tables. We assume that we have a minimum grid size or lookup table size $N$ and that we are provided with lower and upper values for the array $\bar{x}$, namely $x_s$ and $x_f$ respectively. Algorithms CPU1 and CPU2 are single threaded CPU algorithms used to benchmark results. Algorithms FS, DS1, DS2 and DS3 are GPU algorithms constructed such that each lookup table (consisting of $\bar{x}$ and $\bar{P}_x$) is created within a separate GPU thread block.

#### 3.3.1. Naive CPU algorithm (CPU1) and optimised for clustering CPU algorithm (CPU2)

Starting from a single initial pair $\bar{x}(1) = x_s$ and $\bar{P}_x(1) = P(x_s)$, new pairs are added based on a fixed increment of $x$, namely $\Delta x$, where:

$$\Delta x = \frac{x_f - x_s}{N - 1}. \tag{28}$$

Each additional $i$-th pair within $\bar{x}$ and $\bar{P}_x$ is calculated as:

$$\bar{x}(i+1) = \bar{x}(i) + \Delta x, \tag{29}$$
$$\bar{P}_x(i+1) = P(\bar{x}(i) + \Delta x). \tag{30}$$

Step 1: Build two arrays $\bar{x}, \bar{P}_x$ (called $\bar{A}$) of size $N = 12$:

Step 2: Build array $\bar{\zeta}$ of size $N - 1$, locate contiguous error zones:

Step 3: Memory shifting for each of the contiguous error zones:

Step 4: Update vacant values within the refined error zones:

Figure 14: An illustration of algorithm DS3 creating a cumulative probability lookup table where the initial grid size $N = 12$. The shifting of points (step 3) and the update of vacant values (step 4) is conducted sequentially for each contiguous error zone found. Within each error zone however the shifting of points (step 3) and the update of vacant values (step 4) is undertaken by multiple parallel threads.

After each new pair is added, $\bar{\zeta}_i$ shown in (27) is evaluated. If it is found that $\bar{\zeta}_i$ is greater than our chosen interpolation error cap $\zeta^*$, the points $\bar{x}(i+1)$ and $\bar{P}_x(i+1)$ are discarded and replaced with a pair of points based on half the previous increment of $x$, that is:

$$\bar{x}(i+1) = \bar{x}(i) + \frac{\Delta x}{2}, \tag{31}$$
$$\bar{P}_x(i+1) = P\left(\bar{x}(i) + \frac{\Delta x}{2}\right). \tag{32}$$

The process of discarding and replacing pairs ($\bar{x}(i+1)$ and $\bar{P}_x(i+1)$) and halving the increment $\Delta x$ is contin-ued until $\bar{\zeta}_i < \zeta^*$. Subsequently algorithm CPU1 reverts to adding new pairs based on the original increment of $\Delta x$ shown in (28) whereas algorithm CPU2 adds new pairs based on doubling the previously experienced increment of $x$ (with the maximum increment being the original increment of $\Delta x$ shown in (28)). Algorithm CPU2 is thus better able to capture clustering effects (as shown in figure 12b) and consequently has better performance as demonstrated in the results within section 3.4. Additional pairs in both algorithms continue to be added until $\bar{x}(i) \geq x_f$.

### 3.3.2. GPU fixed size algorithm (FS)

This GPU algorithm constructs the arrays $\bar{x}$ and $\bar{P}_x$ of size $N$ based on a fixed increment of $\Delta x$ shown in (28). Each element within these arrays is calculated by a separate parallel thread. The algorithm operates in a fixed grid size $N$ and does not attempt to cap maximum interpolation error $\zeta^*$.

### 3.3.3. GPU dynamic size algorithms (DS)

These GPU algorithms attempt to cap maximum interpolation error $\zeta^*$ and we consider three possible methods of implementation, namely algorithms DS1, DS2 and DS3.

The algorithms initially construct the arrays $\bar{x}$ and $\bar{P}_x$ of size $N$ as shown in algorithm FS. Next an error array $\bar{\zeta}$ of size $N-1$ is constructed as shown in (27), where each element within $\bar{\zeta}$ is created by a separate parallel thread.

*Algorithms DS1 and DS2*

The algorithms use a single thread to loop through the elements $\bar{\zeta}_i$ where $i \in \{1, \cdots, N-1\}$ and are halted at $i$ when $\bar{\zeta}_i > \zeta^*$.

When halted at $i$, values within $\bar{x}$ and $\bar{P}_x$ corresponding to elements $\{i+1, \cdots, N\}$ are transferred or shifted to elements $\{i+2, \cdots, N+1\}$ and values within $\bar{\zeta}$ corresponding to elements $\{i+2, \cdots, N-1\}$ are transferred to elements $\{i+3, \cdots, N\}$. As a result the elements $\bar{x}(i+1)$, $\bar{P}_x(i+1)$, $\bar{\zeta}_i$ and $\bar{\zeta}_{i+1}$ are considered vacant. Algorithm DS1 undertakes this shift using a single thread whereas algorithm DS2 undertakes this shift using multiple threads in parallel.

Additionally, when halted at $i$ (and after the previously stated memory transfer or shift) a single thread updates values for the vacant elements $\bar{x}(i+1)$, $\bar{P}_x(i+1)$, $\bar{\zeta}_i$ and $\bar{\zeta}_{i+1}$. These updated values are calculated based on halving the original increment of $x$, $\Delta x$, as shown in (31) and (32). The algorithms do not increment the halted counter $i$ until $\bar{\zeta}_i < \zeta^*$, in doing so it is possible that multiple memory shifts and updates on $\bar{x}(i+1)$, $\bar{P}_x(i+1)$, $\bar{\zeta}_i$ and $\bar{\zeta}_{i+1}$ are needed. For every such memory shift and update, the increment $\Delta x$ is halved and the array size $N$ is incremented by one.

*Algorithm DS3*

The algorithm uses a single thread to loop through the elements $\bar{\zeta}_i$ where $i \in \{1, \cdots, N-1\}$. During this loop the algorithm locates contiguous regions or error zones where $\{\bar{\zeta}_i, \cdots, \bar{\zeta}_{i+n}\} > \zeta^*$. For each located error zone, multiple parallel threads shift elements within $\bar{x}$, $\bar{P}_x$ and $\bar{\zeta}$ in order to vacate a space that doubles the number of elements situated within an error zone. Therefore, if an error zone consisted of $n$ points, the shift will result in the same error zone now occupying $2n$ points, thereby doubling the error zone's grid refinement. Within this vacant region (of $2n$ points) values within $\bar{x}$, $\bar{P}_x$ and $\bar{\zeta}$ are updated by multiple parallel threads. The algorithm continues this process of locating and refining contiguous regions or error zones until all elements within the error array $\bar{\zeta}$ have interpolation error $\bar{\zeta}_i < \zeta^*$. A visual depiction of this algorithm is given in figure 14.

Critically in terms of performance, the use of contiguous regions or error zones is advantageous due to the clustering of points with higher interpolation error (as shown in figure 12b), whereby clustering reduces the number of zones (which are calculated sequentially) and increases the number of contiguous points in a zone (which are calculated in parallel).

### 3.4. Performance results

Results are generated using the system properties shown in Appendix A. Our results consider the construction of 192 individual lookup tables, this represents a typical problem size (results for smaller problem sizes offered similar performance characteristics as shown in section 4). As stated, within our GPU implementation each lookup array (consisting of $\bar{x}$ and $\bar{P}_x$) is created within a separate GPU thread block.

In table 2 we list algorithm execution times and average lookup table grid sizes $N$ (where the initial grid size is preset to $N = 500$) for various interpolation error cap values $\zeta^*$. Algorithm DS3 strongly outperforms algorithms DS1 and DS2 due to the aforementioned clustering effect captured by algorithm DS3. Algorithm CPU2

| Error cap $\zeta^*$ | None | 5e-4 | 1e-4 | 5e-5 | 1e-5 | 5e-6 | 1e-6 | 5e-7 | 1e-7 |
|---|---|---|---|---|---|---|---|---|---|
| Grid size $N$ | 500 | 503.0 | 543.1 | 585.0 | 785.9 | 959.3 | 1,674 | 2,232 | 4,573 |
| DS1 $t$ (ms) | 2.8 | 30.8 | 249.5 | 507.6 | 1,633.8 | 2,648 | N/A | N/A | N/A |
| DS2 $t$ (ms) | 3.2 | 4.6 | 14.7 | 26.4 | 79.5 | 127.6 | N/A | N/A | N/A |
| DS3 $t$ (ms) | 2.6 | 4.3 | 8.2 | 10.6 | 16.1 | 20.0 | 31.6 | 41.8 | 80.9 |
| CPU1 $t$ (ms) | 114.2 | 118.5 | 134.3 | 156.4 | 278.0 | 406.2 | 950.5 | 1,482 | 3,697 |
| CPU2 $t$ (ms) | 115.5 | 115.5 | 128.6 | 143.1 | 216.4 | 272.2 | 504.8 | 697.2 | 1,469 |
| FS $t$ (ms) | 0.7 | N/A | N/A | N/A | N/A | N/A | N/A | N/A | N/A |
| Speedup | 44.4 | 26.8 | 15.7 | 13.5 | 13.4 | 13.6 | 16.0 | 16.7 | 18.2 |
| DS3 Tpp | 0.005 | 0.009 | 0.015 | 0.018 | 0.020 | 0.021 | 0.019 | 0.019 | 0.018 |
| CPU2 Tpp | 0.231 | 0.230 | 0.237 | 0.245 | 0.275 | 0.284 | 0.302 | 0.312 | 0.321 |

Table 2: Algorithm performance for the generation of 192 cumulative probability lookup tables. Grid size $N$ = average grid size where the initial grid size is preset to $N = 500$. $t$ = execution time (ms). DS1/2/3 = GPU dynamic size algorithms, FS = GPU fixed size algorithm with no error cap, CPU1/2 = CPU algorithms. DS1/DS2 show results limited to an error cap $\zeta^* = 5e-6$. Speedup reported as CPU2 time / DS3 time. Tpp represents the time taken per point and is calculated as Time (ms) / Grid size ($N$). CPU results are based on a single threaded implementation.

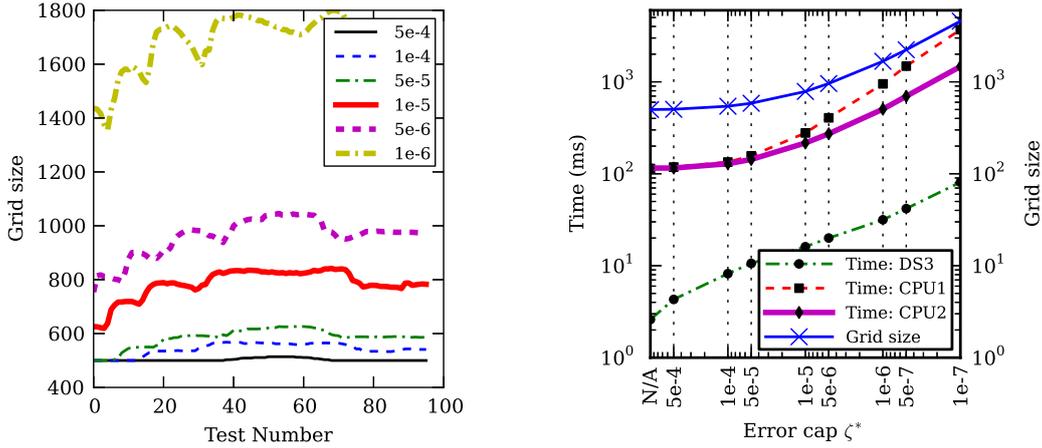

(a) Resultant lookup table grid size $N$ for various error caps $\zeta^*$. All algorithms depicted near identical resultant grid sizes.

(b) Algorithm execution times for various error caps $\zeta^*$. DS3 = GPU dynamic size 3 algorithm. CPU1/2 = CPU algorithms. Grid size = average resultant grid size $N$.

Figure 15: Performance of algorithms generating cumulative probability lookup tables for various interpolation error caps $\zeta^*$.

also outperforms algorithm CPU1 due to the same reason.

In figure 15a we compare the resultant lookup table grid size $N$ for various error caps $\zeta^*$. Within figure 15b, we observe in the higher accuracy region where $\zeta^* = 5 \times 10^{-5}$ to $1 \times 10^{-7}$ that algorithm CPU1 markedly deteriorates as accuracy increases, with calculation time of $O(N^c)$, where $c > 1$. This was due to the extremely heavy discarding of points as is intrinsic within algorithm CPU1. In contrast (within the same higher accuracy region) the performance of algorithms DS3 and CPU2 is based on a superior stable calculation time of $O(N)$. Such stability is evidenced by the consistent Tpp (time taken per point) ratio shown in table 2 and is attributed to the described clustering effect captured by algorithms DS3 and CPU2. As accuracy increases within the stated higher accuracy region, algorithm DS3 maintains a very stable Tpp ratio whereas algorithm CPU2 is shown to experience a small increase in Tpp ratios. This results in a small increase in comparative CPU2 / DS3 speedups as shown in table 2 (the depiction of a more efficient CPU implementation is outside the scope of this paper).

Of further interest within figure 15b is the region where we go from no error cap to the smallest error cap of $\zeta^* = 5 \times 10^{-4}$ which results in a small increase in the average lookup table grid size $N$ (from $N = 500$ to $N = 503$ as shown in table 2). Our CPU implementations have a very low time penalty for the increased grid size due to minimal discarding alongside the very few number of additional points needed within $N$. In contrast algorithm DS3 approximately doubled in execution time due to the additional points being calculated sequentially from the original set of points (despite such points being very few and despite being themselves calculated in parallel).

|  | Coupons | 96 | 48 | 96 | 48 |
|---|---|---|---|---|---|
|  | Valuations | 192 | 96 | 192 | 96 |
| CPU stages | Step 1 (ms) | 20.8 | 11.3 | 42.8 | 20.0 |
|  | Step 2 (ms) | 143.1 | 73.0 | 2072 | 1033 |
|  | Total (ms) | 163.9 | 84.4 | 2115 | 1053 |
|  | $R_c$ error | 3e-7 | 4e-7 | 3e-9 | 3e-9 |
| GPU stages | Step 1 (ms) | 5.9 | 3.3 | 7.9 | 4.2 |
|  | Step 2 (ms) | 10.6 | 5.3 | 109.5 | 58.6 |
|  | Total (ms) | 16.5 | 8.5 | 117.4 | 62.8 |
|  | $R_c$ error | 2e-7 | 3e-7 | 2e-9 | 2e-9 |
| Speedups | Step 1 ($\times$) | 3.54 | 3.46 | 5.41 | 4.78 |
|  | Step 2 ($\times$) | 13.53 | 13.89 | 18.93 | 17.63 |
|  | Total ($\times$) | 9.96 | 9.89 | 18.02 | 16.78 |

Table 3: Overall model performance results. Valuations refer to the number of SABR calibrations (step 1) or the number of created probability lookup tables (step 2). Speedups reported as CPU time / GPU time. $R_c$ error is based on a benchmark value of $R_c$ calculated in section 4.1. CPU results are based on a single threaded implementation.

## 4. Overall empirical error and performance

This section presents an analysis of empirical error propagation and performance results for our derivative pricing model which incorporates both SABR calibration (as described in section 2) and the creation of cumulative probability lookup tables (as described in section 3). Our chosen category of derivatives to be priced is composed of $n$ coupons, where the price of such derivatives is equal to $\sum_{i=1}^{n} PV_c$, where $PV_c$ is the individual and independent price of the $i$-th coupon calculated as:

$$PV_c = R_c \times \text{Not} \times \delta \times \text{DF}, \qquad (33)$$

where $R_c$ is the pricing model calculated coupon rate, Not is the fixed notional amount of the derivative contract reported in units of a currency (e.g. USD), $\delta$ is the time interval upon which $R_c$ acts and DF is a discount factor applied. The magnitude of $R_c \times \text{Not} \times \delta \times \text{DF}$ is typically of $O(10^6$ to $10^9)$, therefore in order for $PV_c$ to be accurate to within 1 unit of a currency (e.g. 1 USD which is a typical choice) the model calculated coupon rate $R_c$ must be accurate to $O(10^{-6}$ to $10^{-9})$.

The coupon rate $R_c$ is calculated through the following sequential steps: 1. SABR calibration, 2. cumulative probability lookup table construction and 3. Monte Carlo simulation. Within our error and performance analysis we focus on steps 1 and 2 only; error and performance analysis of the Monte Carlo simulation in step 3 is outside the scope of this paper. As described in sections 2 and 3, GPU based calculations relating to each SABR calibration (step 1) and each probability lookup table (step 2) are undertaken on separate GPU thread blocks.

*4.1. Error test 1*

Within our first test we fix the accuracy of step 1 at machine precision, such that $\epsilon_{\text{SABR}}$ calculated as the RHS of (7) is of $O(10^{-16})$. Next we calculate a benchmark value of $R_c$ with step 2 interpolation error shown in (27) capped at our lowest feasible level of $\zeta^* = 1 \times 10^{-11}$ (this value resulted in a lookup table size $N$ of $O(10^6)$). Next we calculate a number of test values of $R_c$ with higher interpolation error caps of $\zeta^* > 1 \times 10^{-11}$. The differences between such test values and the benchmark value are reported as errors of $R_c$. Figure 16a depicts error cap values $\zeta^*$ needed to calculate $R_c$ to our target accuracy of $O(10^{-6}$ to $10^{-9})$.

*4.2. Error test 2*

Within our second test we fix the accuracy of step 2 at $\zeta^* = 5 \times 10^{-8}$. Using the benchmark value of $R_c$ calculated in our first error test, we depict magnitudes of $\epsilon_{\text{SABR}}$ needed to calculate $R_c$ to our target accuracy of

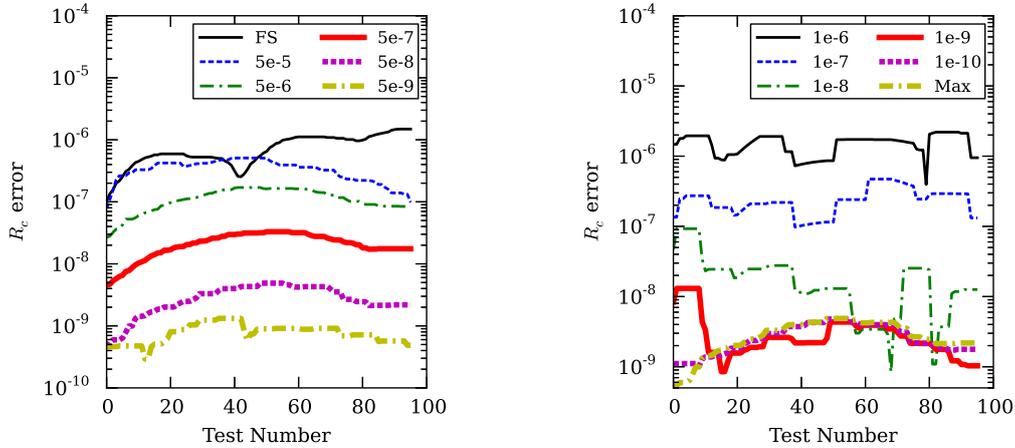

(a) Various interpolation error caps $\zeta^*$. FS = no error cap. The accuracy level $\epsilon_{SABR}$ is fixed at $O(10^{-16})$.

(b) Various SABR accuracy levels $\epsilon_{SABR}$. The interpolation error cap $\zeta^*$ is fixed at 5e-8.

Figure 16: Error in coupon rate $R_c$ for various SABR accuracies $\epsilon_{SABR}$ calculated as the RHS of (7) and for various interpolation error caps $\zeta^*$ based on (27). $R_c$ error is based on a benchmark value of $R_c$ calculated in section 4.1. The required accuracy of $R_c$ depends on parameters within (33) and is typically of $O(10^{-6}$ to $10^{-9})$. Upper error bounds are shown.

$O(10^{-6}$ to $10^{-9})$. Our choice $\zeta^* = 5 \times 10^{-8}$ is motivated by the fact that this value enables $R_c$ to be calculated to our highest target accuracy of $O(10^{-9})$. Since we fix $\zeta^* = 5 \times 10^{-8}$, the best accuracy shown in figure 16b is limited by the bound of $\zeta^* = 5 \times 10^{-8}$ within figure 16a.

*4.3. Timing results*

Within our timing results we utilise two test sets. The first test set uses parameters $\{\epsilon_{SABR} = 1 \times 10^{-7}, \zeta^* = 5 \times 10^{-5}\}$, based on figures 16a and 16b both these parameters result in a predicted $R_c$ error of $O(10^{-7})$. The second test set uses parameters $\{\epsilon_{SABR} = 1 \times 10^{-10}, \zeta^* = 5 \times 10^{-8}\}$, based on figures 16a and 16b both these parameters result in a predicted $R_c$ error of $O(10^{-9})$. Parameters are chosen to reflect the typical desired accuracy of $R_c$ which is of $O(10^{-6}$ to $10^{-9})$.

The two test sets are applied to example derivative contracts having 48 or 96 coupons. Our chosen derivative pricing model is used to value spread derivatives and as a result, for each coupon present, two sets of SABR calibrations and cumulative probability lookup tables are needed. Results are presented in table 3, where $R_c$ errors are shown to match our stated predictions. The total performance speedup (single threaded CPU time / GPU time) is from 10 to 20 ×. Reiterating our earlier performance analysis (in sections 2 and 3) we note that the GPU implementation offers superior comparative performance with more accurate error bounds.

## 5. Conclusions

This paper has presented novel implementations of iterative algorithms common within derivative pricing models, namely SABR calibration and the creation of cumulative probability lookup tables. The proposed SABR calibration method is well suited to parallel architectures such as GPUs where use of the relative value (RV) method with local mesh refinement offered fast convergence. The SABR calibration method was also shown to be correct for SABR models that are consistent with a number of stated properties. Further work may consider a direct mathematical analysis of the underlying SABR equations to determine correctness of the stated properties and may also consider instances where calibration is based on more than three market instruments. The creation of cumulative probability lookup tables through the proposed dynamic size 3 algorithm (DS3) is also well suited to parallel architectures due to the use of parallelism to capture clustering effects that arise within certain probability distributions.

**Acknowledgements**

I would like to thank Ian Eames, Graham Barrett and anonymous reviewers for their helpful comments. The study was supported by the UK PhD Centre in Financial Computing, as part of the Research Councils UK Digital Economy Programme, and BNP Paribas.

**Appendix A. System Properties**

Computational results are based on the following: CPU results are based on an Intel Xeon L5640 processor with results reported from a single thread process using C++ code compiled using the Microsoft Visual Studio 2010 compiler. GPU results are based on an NVIDIA Tesla M2070 (Fermi generation) processor using CUDA C code compiled under the NVIDIA CUDA 4.2 runtime API. Both CPU and GPU implementations are based on calculations using double precision floating point variables.